\documentclass[aps,twocolumn,amsfonts,amssymb,amsmath]{revtex4}
\usepackage{epsfig, graphicx, color, amsmath, amssymb}
\def\ket #1{\vert #1\rangle}

\begin{document}

\author{Ben W. Reichardt}
\email{breic@cs.berkeley.edu}
\thanks{Research supported in part by NSF ITR Grant CCR-0121555.}
\affiliation{UC Berkeley}

\title{Improved ancilla preparation scheme increases fault-tolerant threshold}

\begin{abstract}
We demonstrate an improved concatenated encoded ancilla preparation procedure.  Simulations show that this procedure significantly increases the error threshold beneath which arbitrarily long quantum computations are possible.
\end{abstract}
\maketitle

\section{Introduction}

A quantum computer will need to be engineered to reduce the effect of errors to acceptable levels.
There are a number of techniques that can be applied together to achieve fault tolerance, including decoherence free subsystems, dynamical decoupling, and error-correcting codes \cite{ByrdWuLidar}.
Here we concentrate on engineering an efficient code correction procedure.

For noise below a certain threshold error rate, multiply concatenated coding allows the effective logical error rate to be reduced so arbitrarily long computations can be performed.  This threshold, which depends on the error model, is a good benchmark for an error correction scheme.  It is a fundamental theoretical and engineering task to devise schemes to increase the threshold.  Estimates of the threshold with previous schemes have varied widely.  Aharonov and Ben-Or \cite{AharonovBenOr99} estimate a noise threshold of $10^{-6}$ for a scheme with no classical computation.  Using measurement together with fast and accurate classical computation increases the threshold.  Gottesman and Preskill \cite{Preskill97} estimate a threshold between $10^{-4}$ and $6 \times 10^{-4}$.  Zalka \cite{Zalka97} gives a scheme for which he optimistically estimates a gate noise threshold of $10^{-3}$.  Steane \cite{Steane03} recently gave a further optimized error correction procedure which increases the noise threshold to almost $3 \times 10^{-3}$.

In this paper, we give new optimizations to maximize the efficiency of the Hamming $[[7,1,3]]$ quantum error correcting code.  We design an encoded ancilla preparation procedure which increases the fault-tolerant threshold to almost $9 \times 10^{-3}$ (in a particular error model),
nearly a factor of three improvement.  At a noise rate of $3 \times 10^{-3}$, the new procedure increases efficiency by two orders of magnitude.  Efficient classical simulations of the quantum error process demonstrate these results.

Besides exhibiting this one procedure, a broader goal of this paper is to show how carefully engineering the error correction procedure has the potential to dramatically increase its efficiency.

In Section~\ref{s:CSScodes} we focus on CSS codes, and summarize the error correction and ancilla preparation methods of Steane developed incrementally over a series of papers.

In Section~\ref{s:rejectancilla} we present our improved ancilla preparation scheme, which increases the number of checks an ancilla must pass in order to be used in error correction.

In Section~\ref{s:analysis}, we use simulations to analyze the performance of the improved ancilla preparation scheme together with a slightly modified error correction scheme.  We also analyze the efficiency of the new scheme -- how much time, or how much parallelism, it takes to prepare an ancilla.

In Section~\ref{s:ideal}, we ask by how much could similar optimizations increase the error correction threshold.  We simulate idealized models to speculatively upper bound the increases possible using optimizations of this type.

\section{Error correction procedure} \label{s:CSScodes}

\subsection{Syndrome extraction using an encoded ancilla}

\begin{figure}
\includegraphics*[bb=53 592 599 759,scale=.35,clip=true]{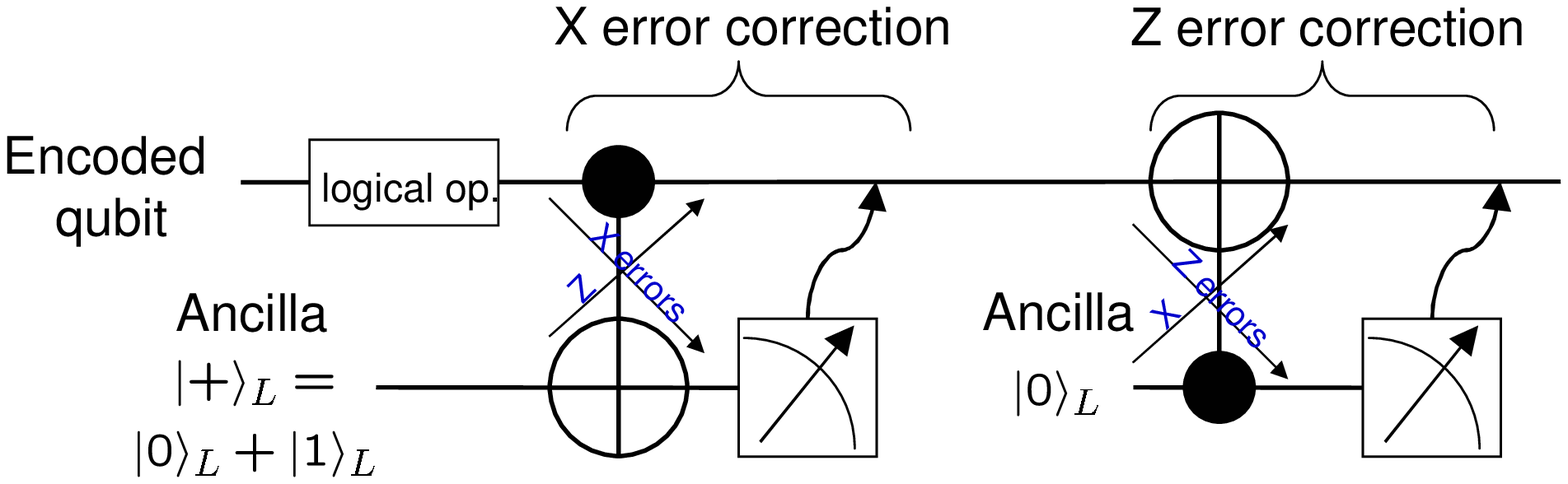}
\caption{}\label{f:errorcorrection}
\end{figure}

A conceptual diagram of the error correction procedure is shown in Fig.~\ref{f:errorcorrection}.  Here, each wire represents a logical bit, encoded perhaps recursively with the Hamming $[[7,1,3]]$ code.  The logical CNOT gates correspond to physical CNOT gates acting transversely.

After every fault-tolerant logical operation, we correct for errors -- first $X$ bit flip errors, then $Z$ phase flip errors.  To correct for $X$ errors, we prepare an ancilla in the encoded $\ket{+}_L \equiv \tfrac{1}{\sqrt{2}}(\ket{0}_L+\ket{1}_L)$ state (logical $X$ $+\!1$ eigenstate).  We then apply a transverse CNOT gate from data to ancilla, moving $X$ errors from the data to the ancilla.  We measure each bit of the ancilla in the $Z$ eigenbasis.  The obtained syndrome allows us to locate the $X$ errors just as for the classical Hamming code.

Notice that 1. $Z$ errors from the ancilla move to the data, and 2. $X$ errors in the ancilla will lead to an incorrect syndrome diagnosis.  Steane gives an error correction procedure tailored to address these two problems \cite{Steane03}.  To address the first problem, Steane takes care to prepare the ancilla with as few $Z$ errors as possible.  Then to avoid basing a correction on an incorrect syndrome, he uses repeated syndrome extraction (not shown).  If the first syndrome extracted is nontrivial, he extracts two more syndromes and only applies a correction if at least two syndromes agree.  If no two syndromes agree, he does not try to extract more syndromes -- after all, $Z$ errors may be accumulating in the data.  Instead, in the next round of $X$ error correction he will extract one more syndrome and (if it is nontrivial) he looks for agreement with one of the previous syndromes before applying a correction.

$Z$ error correction is entirely symmetrical.  Full details of the procedure can be found in \cite{Steane03}.  For example, if the data block is the control of a logical CNOT gate, Steane corrects for $Z$ errors, then $X$ errors before the logical gate -- this makes it less likely for $X$ errors to propagate across the gate from one data block to another.

\subsection{Ancilla preparation} \label{s:CSSancilla}

It remains to describe how Steane prepares the encoded ancilla $\ket{+}_L$.  The basic idea, described in \cite{Steane02} for general CSS codes, is to maximize parallelism while generating the ancilla, and then to verify the $X$ stabilizers to minimize $Z$ errors.  Moreover, the verification process is designed so that $Z$ errors of weight $k$ occur with probability at most order $\gamma^k$ (despite correlations introduced in the generation step).

\begin{figure}
\includegraphics*[bb=51 249 762 699,scale=.30]{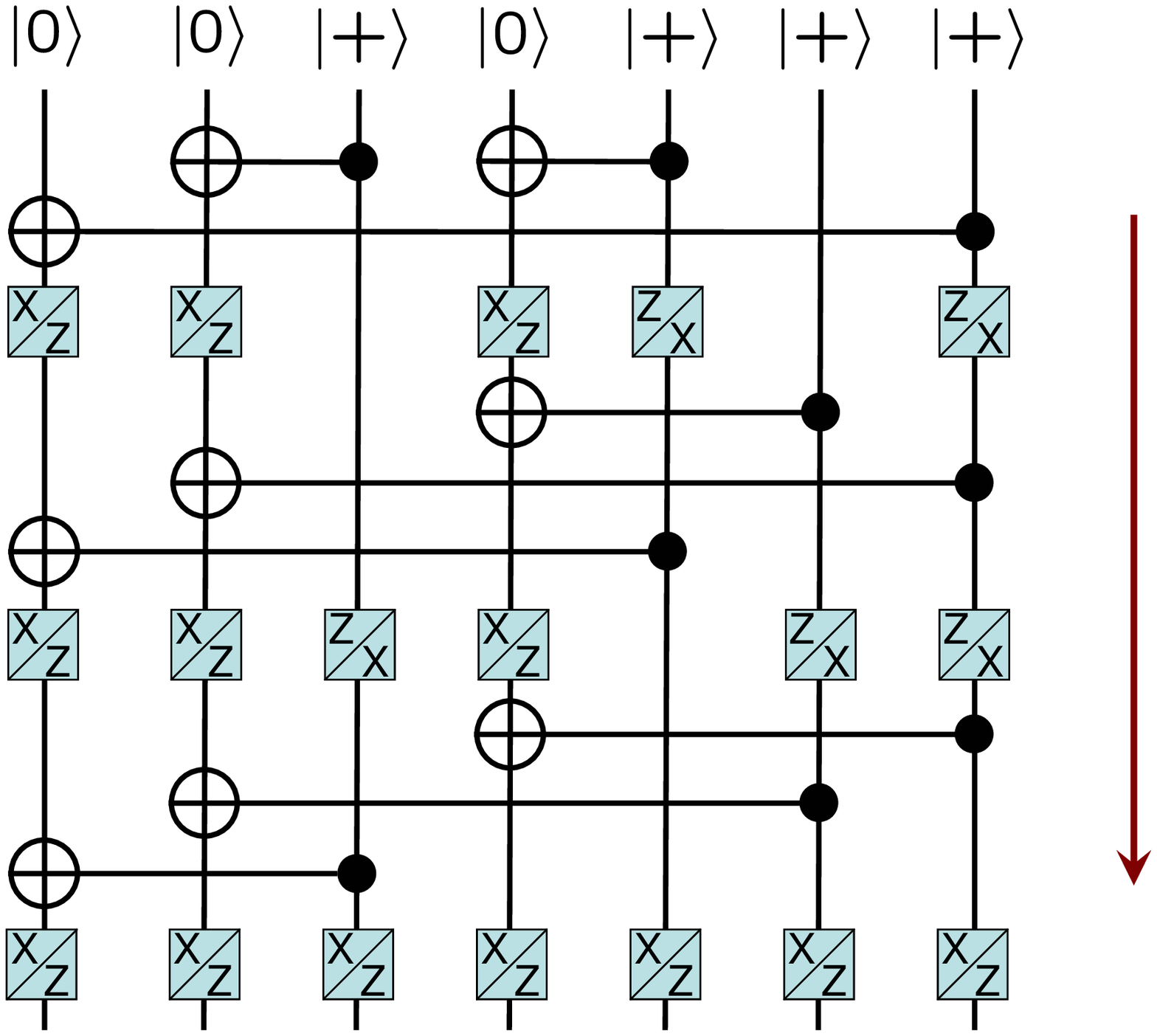}
\caption{}\label{f:ancillapreparation}\bigskip

\includegraphics*[bb=51 247 762 769,scale=.30]{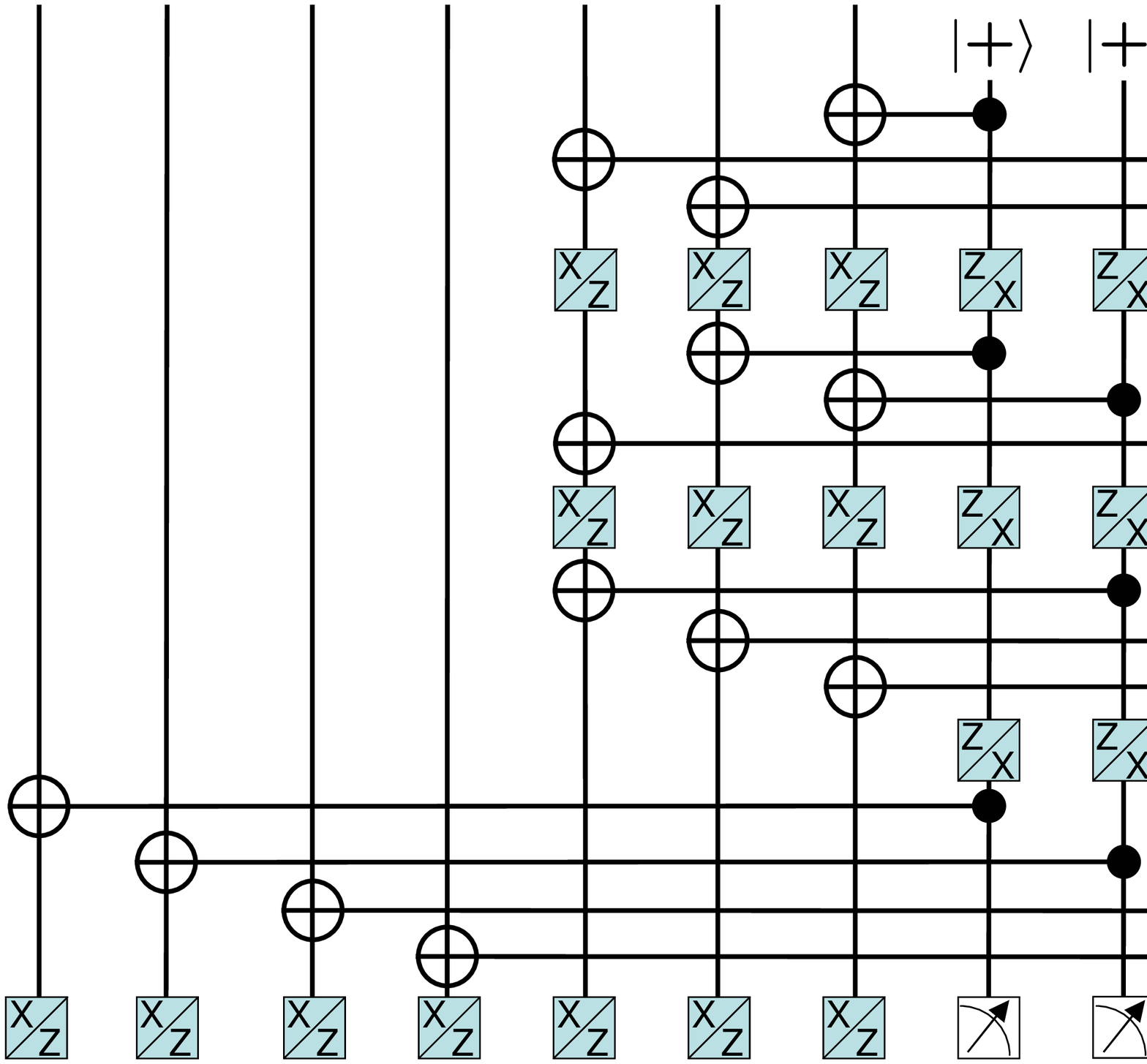}
\caption{}\label{f:ancillaverification}
\end{figure}

Steane's procedure to prepare $\ket{+}_L$ with minimal $Z$ errors for the Hamming code is shown in Fig.~\ref{f:ancillapreparation}.  The boxes indicate error correction -- $X$ then $Z$, or $Z$ then $X$ -- at the next lower level of concatenation, if applicable.  The verification of the $X$ stabilizers is shown in Fig.~\ref{f:ancillaverification}.

Many ancillas are generated and verified in parallel.  Only those passing the verification are used for error correction.

\section{Improved ancilla preparation and error correction} \label{s:rejectancilla}

We implement a single optimization to Steane's ancilla preparation procedure.  We prepare the ancilla as in Fig.~\ref{f:ancillapreparation}.  However, every error correction step is replaced by an error \emph{detection} step.  If a nonzero syndrome is detected, all 49 bits of the ancilla are discarded, and we start anew.  This changes the probability of a logical error from a second-order to a third-order event.  For efficiency, we do not bother with the logical checks of Fig.~\ref{f:ancillaverification}.  (The error detection uses 7 bit ancillas prepared as by Steane.)

In order to take advantage of the improved ancillas, we also simplify Steane's error correction procedure.
We extract only a single syndrome, and correct for it considering the concatenated code as a $[[49,1,9]]$ code (i.e., correcting up to four errors).
We do not extract multiple syndromes to check for mistakes.  We do not correct at the 7-bit code level.

\section{Analysis} \label{s:analysis}

\subsection{Error model}

We work with a simple depolarization error model which lends itself to experimentation.  When an error occurs on a bit, that bit is depolarized -- replaced by the uniform density matrix.  To implement this, we record with equal probabilities $1/4$ either $I$ (no error), $X$, $Y$ or $Z$.

There are four basic operations: preparation of fresh bits (as either $\ket{0}$ or $\ket{+}$), single bit gates, two bit CNOT gates, and destructive measurement.  Each of these operations has probability $\gamma$ of failure, in which case the affected bits are depolarized.  In the case of a two bit gate failure, both bits are depolarized.

We do not consider memory errors (and therefore available gate parallelism is irrelevant).  We also do not consider the physical location of bits within the computer -- so CNOT gates can couple arbitrary bits, not just nearest neighbors.  Our simpler error model facilitates experimentation with different error correction procedures.  By developing an understanding of this simple model, we are better prepared for optimizing error correction for complicated quantum computer engineering tasks, for example a 3-dimensional lattice ion trap computer with lower-level encoding (like a DFS or code against qubit loss \cite{Vala}).

Steane's error correction procedure \cite{Steane03} was developed in a model where memory errors might be significant; he ran simulations with a memory error rate from $\gamma/100$ to $\gamma$.  For a more fair comparison, we evaluated several small optimizations to his procedure effective in the low memory error rate regime.  Steane extracts one syndrome, aborting error correction if that syndrome is trivial.  Otherwise, he extracts two more syndromes, and looks for two agreeing syndromes.  We modify this to always extract one syndrome at a time, and to always abort after retrieving a trivial syndrome -- i.e., if the first two syndromes agree or if the second syndrome is trivial, don't extract a third one.

\subsection{Simulation method}

We largely follow the simulation method and statistical procedures as described by Steane \cite{Steane03}.  The simulator is fast and efficiently scalable since for each bit it only needs to store either $I$, $X$, $Y$ or $Z$.  Although ancilla states are stabilizer states, there is no need to keep track of the stabilizer with a simulator like CHP \cite{AaronsonGottesman} because the stabilizer is always well known.  Even though data blocks may have arbitrary encoded states, there is no need to keep track of the exact, exponentially large quantum state with a simulator like QuIDDPro/D \cite{ViamontesMarkovHayes}, because we only care about the error correction part of the circuit.

To increase the credibility of Steane's simulation method, we additionally track errors between 49-bit data blocks.  This is important because even blocks which have just been corrected still have some errors.  Every round we apply a transverse CNOT gate to or from another data block.  We then correct $X$ and $Z$ errors in a random order which determines whether the block will be the control or target of the next round's transverse CNOT gate.  To implement this, if a block has not crashed we save it in a list for later use.  As a detail, we only save blocks after the third round, to allow the distribution of errors to converge.

\subsection{Results}

\begin{figure}
\includegraphics*[bb=41 322 749 766,scale=.33]{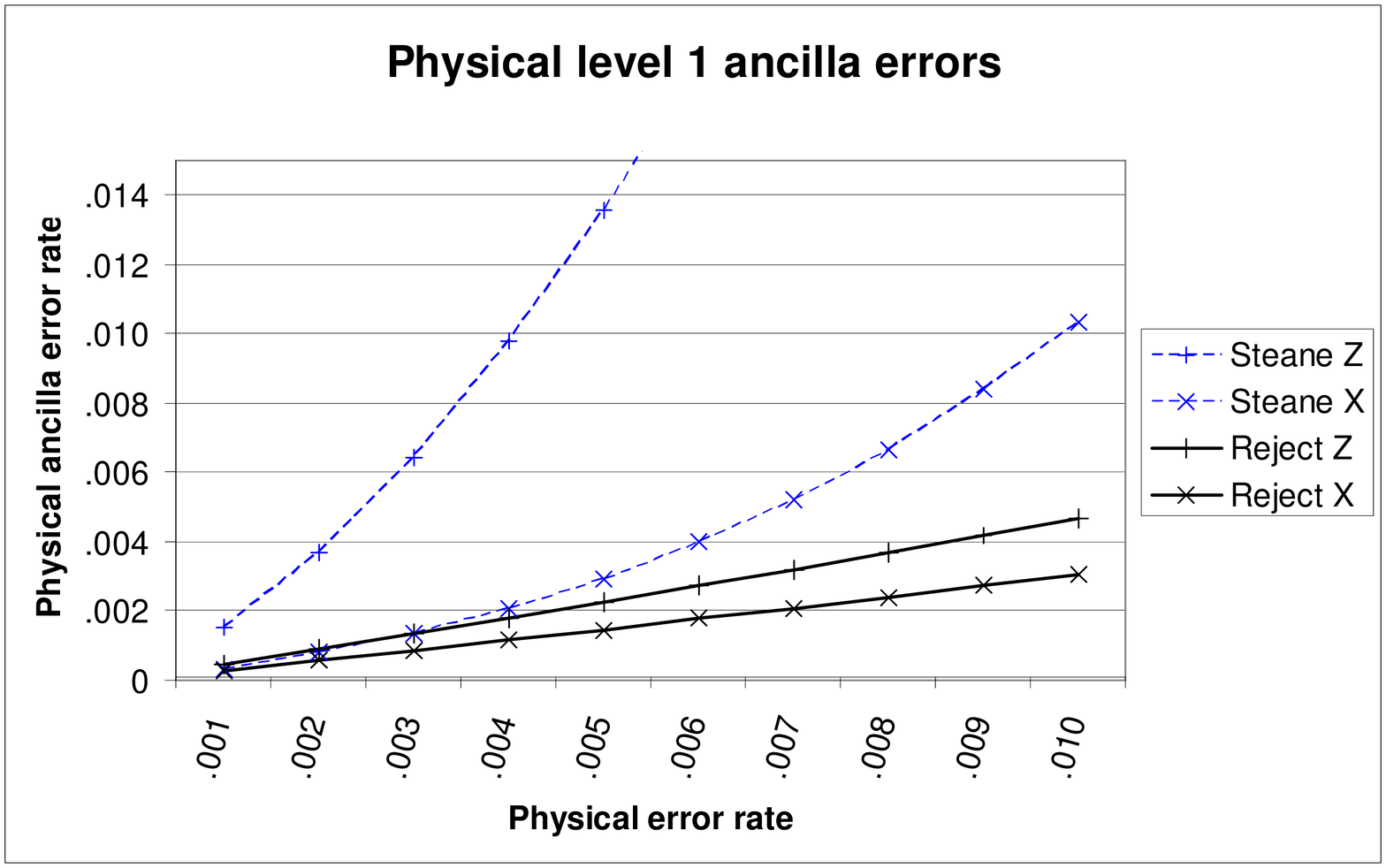}
\caption{}\label{f:physicalancillaerrors}
\end{figure}

\begin{figure}
\includegraphics*[bb=41 322 749 766,scale=.33]{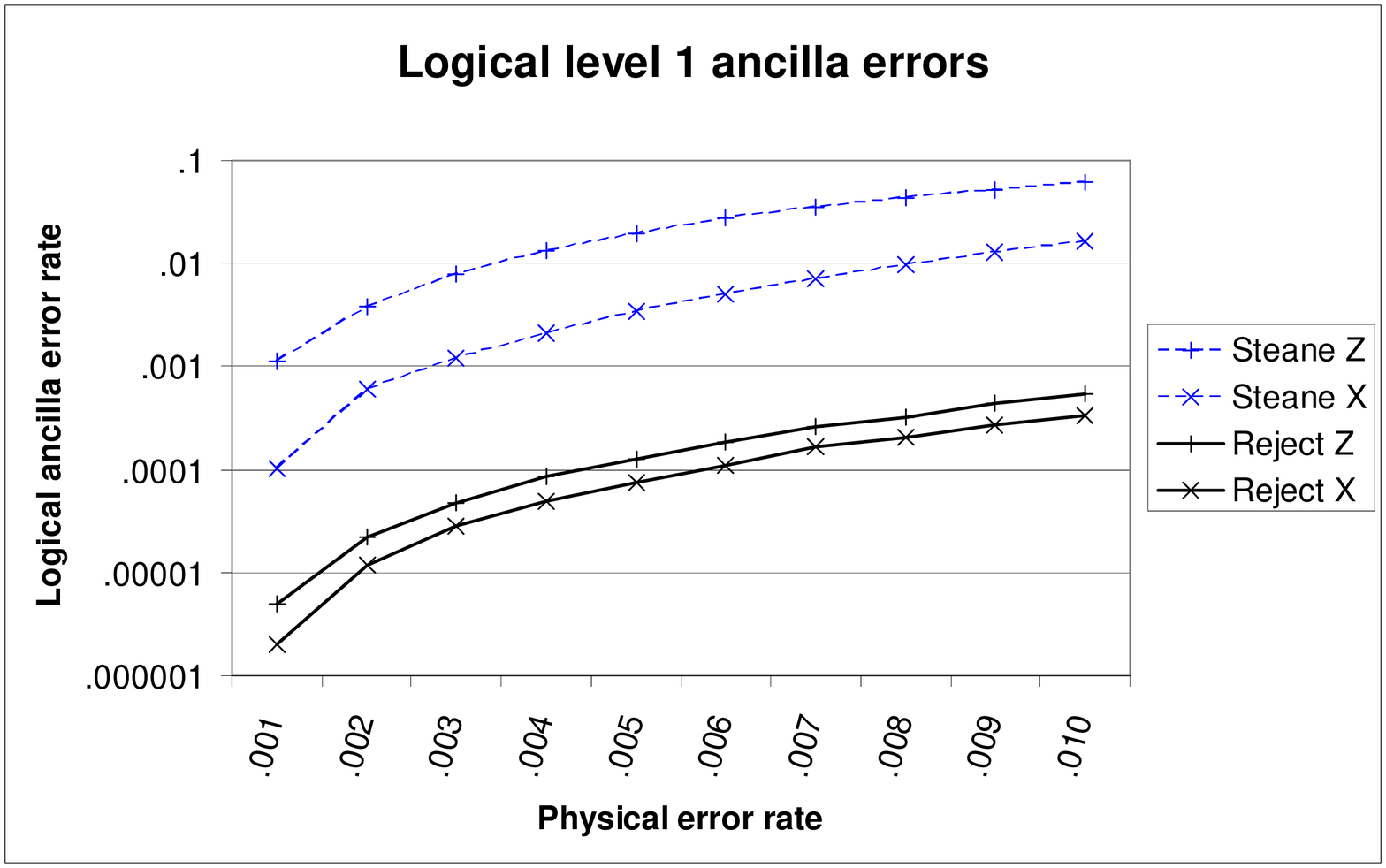}
\caption{}\label{f:logicalancillaerrors}\bigskip

\includegraphics*[bb=41 322 749 766,scale=.33]{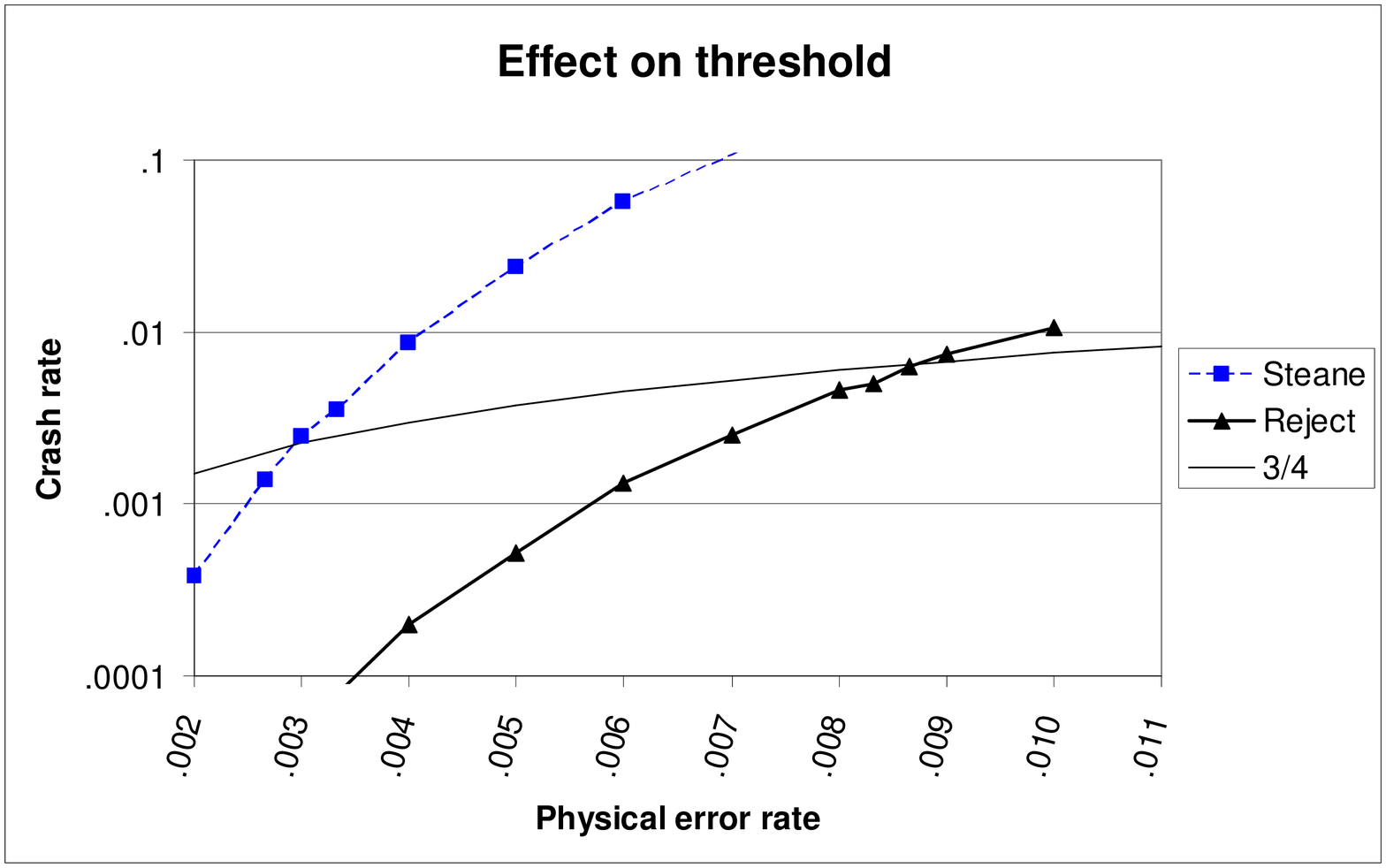}
\caption{}\label{f:effectonthreshold}
\end{figure}

Figure~\ref{f:physicalancillaerrors} shows that the physical error rate is reduced slightly.  This slight reduction is largely because of the dramatic reduction in the logical error rate (2 or 3 bit physical errors within a block) shown in Fig.~\ref{f:logicalancillaerrors}.

Figure~\ref{f:effectonthreshold} shows the effect of the new error correction procedure on the crash rate compared to Steane's procedure.  (A crash is defined as a set of physical errors which perfect error correction would project to a logical error.)  To read off an approximate threshold from the figure, we take the intersection of the crash rate curve with the line of slope $3/4$ (the rate at which $X$, $Y$ or $Z$ physical errors occur is $3/4$ the depolarization rate).  This is only an approximate result, since the error model at the next higher level is not the same as the physical depolarization error model.  In particular $X$ and $Z$ logical errors are less correlated than $X$ and $Z$ physical errors.

\begin{figure}
\includegraphics*[bb=41 322 749 766,scale=.33]{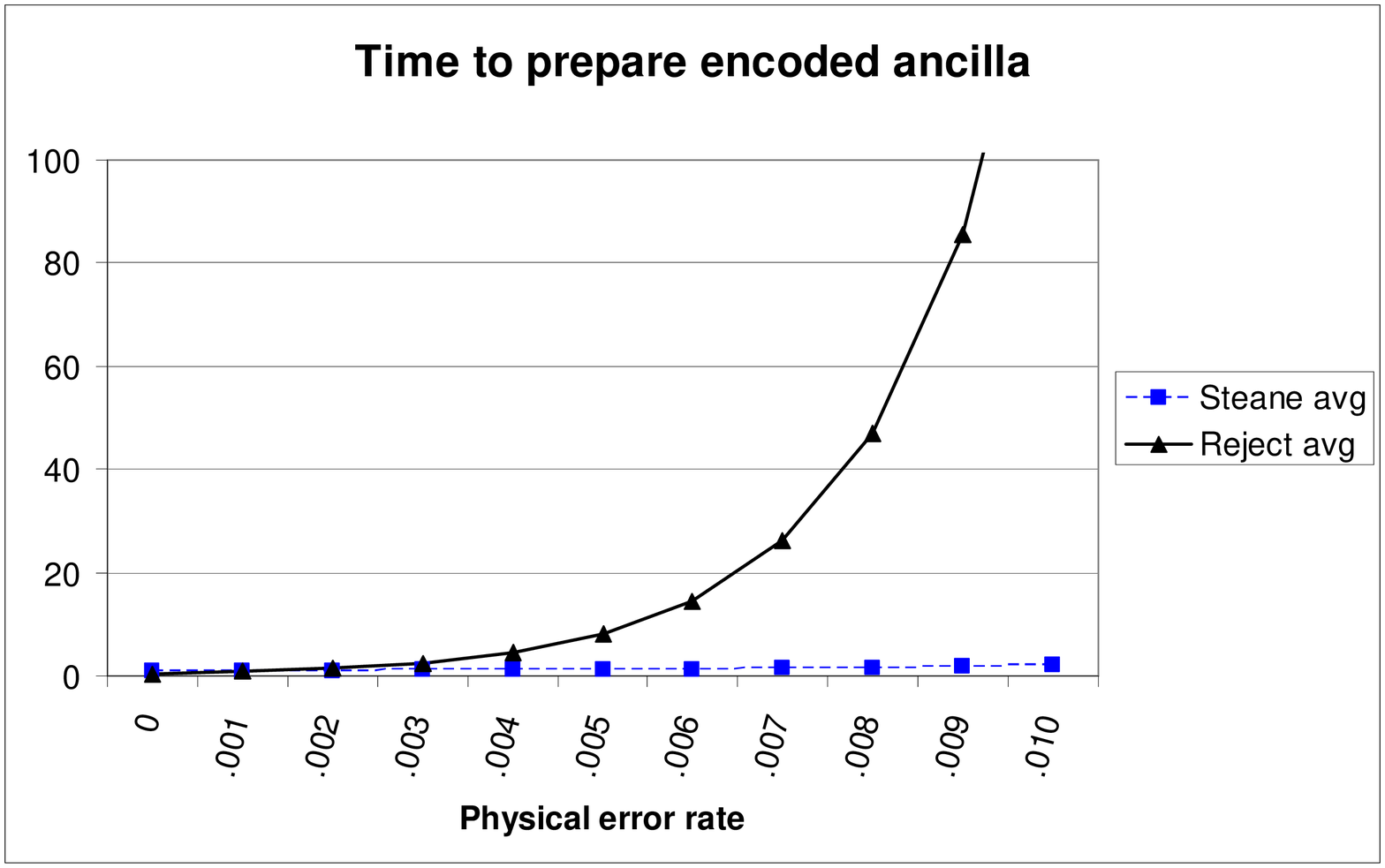}
\caption{}\label{f:timetoprepareencodedancilla}
\end{figure}

There is a loss of efficiency at higher error rates, but at rates at or below the old threshold, the efficiency is actually just as good or better.  See Fig.~\ref{f:timetoprepareencodedancilla}, where preparing a Steane ancilla at zero error rate is defined to take one unit of time.  Since the error rate drops rapidly with concatenation below the threshold, the efficiency loss will show up in practice only at the lowest concatenation levels.

Further optimization could reduce this overhead.  Its exponential dependence on the error rate makes it sensitive to small changes to which the threshold itself is insensitive.  For example, we found that modifying the 7-bit ancilla verification to only check only three of the four relevant stabilizers -- an idea from \cite{Zalka97} -- reduced the overhead by about a factor of four at 1\% error rate.

However, the poor efficiency at high error rates probably rules out some generalizations of this ancilla preparation scheme.  For example, we expect it will be impractical to apply the scheme to the 343 bit code gotten by concatenating the Hamming code twice, or to the $[[23,1,7]]$ Golay code concatenated with itself.

\section{Optimized ancilla preparation} \label{s:ideal}

How much more can optimizations of this type gain us?  We ran the same simulations with a 49-bit ancilla perfectly prepared, except each bit was subject to a single possibility of error, then each block was checked for X and Z errors.  If an error was detected, the entire ancilla was rejected.  For this ideal ancilla, logical errors are much less correlated than in our preparation procedure.

The simulations showed almost no improvement whatsoever in the threshold.  Our ancilla preparation procedure is already quite well optimized, and the crashes that still occur cannot generally be blamed on a faulty ancilla.  Hence further optimizations of the error correction procedure for the $[[49,1,9]]$ code will need to do more than improve ancilla preparation.

We are not able to simulate rejection ancilla preparation for the Golay code concatenated with itself; too many ancillas must be thrown away for every good ancilla.  (A crude estimate puts the overhead at ~$10^{50}$, although optimizations can reduce this enormously.)  We instead simulated preparation of ideal ancillas as above for the Hamming code, with the expectation that results would be close to the actual ancilla preparation procedure.  We found that the threshold increases to about $1\%$.  While $10^{50}$ is impractically large, it is still a constant.  Theoretically, it is of interest how much higher the threshold can be increased by using different codes, more concatenation, or improved syndrome interpretation.

\section{Conclusion}

For quantum computers to become a reality, highly efficient coding against errors is essential.  We have proposed an optimized encoded ancilla preparation scheme for the concatenated Hamming code.  Simulations show that the improved scheme decreases the crash rate by almost two orders of magnitude, and increases the fault-tolerant threshold significantly.  There is however a loss of efficiency at higher error rates.

Very recent independent work by Knill \cite{Knill04} uses a similar idea to increase the threshold.  Further research is needed to make these schemes more practical: still more efficient, and applicable to more general error models.



\end{document}